\documentclass[prb,aps,floats,amssymb,showkeys,showpacs,preprint,
superscriptaddress,tightenlines]{revtex4}
\usepackage{graphicx}
\usepackage{epsfig,bm}
\begin{document}
\preprint{Bicocca-FT-xx-yy  June 2007}

\title{ Further extensions of the high-temperature expansions for the \\  
  two-dimensional classical XY model on the triangular\\
 and the square lattices\\
 }
\author{P. Butera\cite{pb}}
\affiliation
{Istituto Nazionale di Fisica Nucleare \\
Sezione di Milano-Bicocca\\
 3 Piazza della Scienza, 20126 Milano, Italy}

\author{M. Pernici\cite{mp}}
\affiliation
{Istituto Nazionale di Fisica Nucleare \\
Sezione di Milano\\
 16 Via Celoria, 20133 Milano, Italy}

\date{\today}
\begin{abstract}
The high-temperature expansions of the spin-spin correlation function
 of the two-dimensional classical XY (planar rotator) model are 
 extended by two terms, from order 24 through order 26, in the case of
 the  square lattice, and by five terms, from order 15 through order 20,
 in the case of the triangular lattice.  The data are analyzed to improve
 the current estimates of the critical parameters of the models.
We determine $\beta_c=0.5599(7)$ and $\sigma=0.499(5)$ in the case of  
 the square lattice. For the triangular lattice case,  we estimate 
 $\beta_c=0.3412(4)$ and $\sigma=0.500(5)$. 
\end{abstract}

\maketitle

We have recently extended\cite{bper}, from the 21th to the 24th order,
 the high-temperature (HT) expansions for the spin-spin correlation
 function of the classical two-dimensional XY
 Berezinskii-Kosterlitz-Thouless(BKT) lattice model\cite{BKT} with
 nearest-neighbor interactions on the square({\it sq}) lattice and
 briefly updated the series analysis for the susceptibility and the
 second-moment correlation-length.  Here we present a further
 extension of the HT expansions from the 24th order through the 26th
 order for the {\it sq} lattice and from the 15th order to the 20th
 order in the case of the triangular({\it tr}) lattice.  In our
 derivation of the series coefficients, we have used a non-graphical
 recursive algorithm\cite{BC,bct} based on the Schwinger-Dyson
 equations for the spin-spin correlations of the XY-model.  The
 necessary computations were performed using a few nodes of a pc
 cluster for an equivalent single-processor CPU-time of seven weeks in
 the case of the {\it tr} lattice and a CPU-time ten times as long in
 the case of the {\it sq} lattice. Arguing as in
 Ref.[\onlinecite{guttfc}] that the ``{\it effective length}'' (namely
 the ability to provide accurate numerical information) of the HT
 expansions of a given system on different lattices, is proportional
 to the computer time used in the series generation, we expect that
 the ``{\it effective length}'' of our expansions for the {\it tr}
 lattice, is only slightly smaller than that of the {\it sq} lattice
 series and moreover that their convergence will be faster, because
 the {\it tr} lattice is closely-packed.

Let us recall that 
the lattice XY spin model  with nearest-neighbor interactions is described 
by the    Hamiltonian

\begin{equation}
H\{ v \} = - 2{J}  \sum_{<nn>} 
\vec v({\vec r}) \cdot \vec v({\vec r}\;')
\label{hamilt}
\end{equation}
 where $\vec v({\vec r})$ 
 is a two-component classical unit vector  at the  site ${\vec r}$ and the sum
 extends to the nearest-neighbor sites  of the lattice.

We have calculated the spin-spin correlation function
\begin{equation}
C(\vec 0, \vec x;\beta)= <\vec v(\vec 0) \cdot \vec v(\vec x)>,
\label{corfun}
\end{equation}  
 as series expansion in the variable $\beta= 2J/kT$, with $T$ the
temperature and $k$ the Boltzmann constant, for all values of $\vec x$
for which the HT expansion coefficients are non-trivial within the
maximum order reached. In terms of these quantities, we can form the
expansions of the $l$-th order spherical moments of the correlation
function:
\begin{equation}
{\rm m^{(l)}(\beta)} = \sum_{\vec x  }|\vec x|^l 
< \vec v(\vec 0) \cdot  \vec v(\vec x)>
\label{sfermom}
\end{equation}

\noindent and, in particular, of the reduced ferromagnetic 
susceptibility $\chi(\beta)={\rm m^{(0)}}(\beta)$.

The  second-moment correlation length is defined, as usual,
 in terms of $\rm {m^{(2)}}(\beta)$ and $\chi(\beta )$: 
\begin{equation}
\xi^2(\beta) ={\rm m^{(2)}}(\beta)/4\chi(\beta).
\label{corleng}
\end{equation}

 After completing our computations, we have been kindly informed by
 H. Arisue\cite{aris} that, in the particular case of the {\it sq}
 lattice, he has recently obtained, using the ``finite lattice'' method,
 remarkably longer HT expansions: through the 38th order for the
 nearest-neighbor spin-spin correlation $C(0,0;1,0;\beta)$ and through
 the 34th order for the susceptibility and the second and fourth
 moment of the correlation function.

 When also these further extended data will be published, the
 agreement, through their common extent, between both sets of results
 independently obtained by completely different methods, will be a
 significant check of correctness in consideration of the notoriously
 intricate nature of the high-order HT computations.

In order to determine the critical parameters of the models, the HT
expansions of the above defined quantities should be confronted to the
following main predictions\cite{BKT,Kenna} of the BKT
renormalization-group analysis of the XY system.

As $\beta \rightarrow \beta_{c}$, the divergence of the correlation
 length is expected to be dominated by the characteristic singularity
\begin{equation}
\xi(\beta) \sim \xi_{as}(\beta)= \exp(b \tau^{-\sigma})[1+O(\tau)]
\label{xias}
\end{equation}

where $\tau=1-\beta/\beta_{c}$ and the universal exponent $\sigma$ is 
expected to take the value $\sigma=1/2$, 
while $b$ is a non-universal (and thus lattice dependent)  positive constant.

The critical behavior of
 the susceptibility is predicted to be 
\begin{equation}
\chi(\beta) \sim \xi^{2-\eta}_{as}(\beta)=
 \exp \Big( (2-\eta)b \tau^{-\sigma} \Big)[1+O(\tau)]
\label{chias}
\end{equation}
The parameter $\eta$ represents the correlation-function exponent and
is predicted to take the value $\eta= 1/4$.

At the critical inverse temperature $\beta=\beta_{c}$, the asymptotic
behavior of the two-spin correlation function as 
$|\vec x |= r\rightarrow \infty$ is expected\cite{Amit} to be
\begin{equation} 
<\vec v(\vec 0) \cdot \vec v(\vec x)> \sim \frac 
{({\rm ln}r)^{2\theta}} {r^{\eta}}
[1+O(\frac{{\rm ln}{\rm ln}r} {{\rm ln}r})]
\label{coras}
\end{equation}

 The value $\theta = 1/16$ is predicted for the second universal
 exponent characterizing the critical correlation function.

Following a recent renormalization-group analysis\cite{Bal} of the
two-dimensional $O(2)$ nonlinear $\sigma$-model, which should belong
to the same universality class as the XY model, we shall assume that
the critical behavior of $\chi(\beta)$ does not contain singular
multiplicative corrections by powers of $\rm ln(\xi)$ (or equivalently
singular multiplicative corrections by powers of $\tau$).  The possible
existence of such corrections has long been numerically
investigated\cite{Kenna,Irv, Has,pisa} with conflicting and
essentially inconclusive results. As we will indicate below, our
series analysis is completely consistent with the conclusions of
Ref.[\onlinecite{Bal}].

Let us now come to our results for the  {\it sq} lattice: 
we have added the two terms

\begin{equation}
 \chi(\beta)=...+\frac{376988970189597090587}{384296140800}\beta^{25}
+\frac{62337378385915430773643}{31384184832000}\beta^{26}+...
\end{equation}

 to the expansion of the susceptibility.

  To the expansion of the  second moment
of the correlation function, we have added the terms 
\begin{equation}
\rm m^{(2)}(\beta)=
...+\frac{5412508223507386985733313 }{13450364928000}\beta^{25}+
 \frac{7139182711315236460182251}{7846046208000 }\beta^{26}+...
\end{equation}

  The lower-order coefficients of the  {\it sq}-lattice 
expansions of $\chi(\beta)$ and 
 of $\rm m^{(2)}(\beta)$ have been recently
tabulated in Ref.[\onlinecite{bper}] and therefore they will not be
reproduced here.

 On the contrary, we have listed in Table \ref{tabtria} also the
coefficients already known\cite{bct,pisa} through 15th order in the
case of the {\it tr} lattice, in addition to the five coefficients
that we have recently calculated for the nearest-neighbor spin-spin
correlation, the susceptibility and the second moment of the
correlation function.

Let us now update the analysis of Refs.[\onlinecite{bper,bct}] for the
{\it sq} lattice expansions by including all coefficients that we have
so far derived.  The expected form eq. (\ref{xias}) of the critical
singularity suggests that the critical parameters should be
conveniently obtained by an
inhomogeneous-differential-approximant\cite{Gutt} (DA) study of the
location and the exponent of the leading singularity of the quantity
$\rm ln^2(\chi)$.  This conclusion is supported also by a simple
comparison of the distribution of the singularities on the real
$\beta$-axis for the DAs of $\chi(\beta)$ and of $\rm ln^2(\chi)$
which, independently of theoretical prejudice,
 suggests that the analytic structure of the latter form is more
suitable to a study by DAs.  Notice that the choice of this particular
function of $\chi(\beta)$ does not imply any biasing of $\sigma$, as
long as we do not make any selection of the singularities of the DAs
for $\rm ln^2(\chi)$ on the basis of their exponents.  In our previous
note\cite{bper}, we had already observed that a possibly more accurate
analysis might be based on the obvious remark that, near the critical
point, by eq. (\ref{chias}), we have to expect ${\rm ln}(\chi)
=c^{sq}_1/\tau^{\sigma}+c^{sq}_2+..$.  Assuming $\sigma \simeq 1/2$, a
simple fit to the asymptotic form of ${\rm ln}(\chi)$, can determine
$c^{sq}_2 \simeq -1.5$. We are then led to study also the simple
generalization of the function $\rm ln^2(\chi)$, defined by
$L(a,\chi)=( a+ {\rm ln}(\chi))^{2}$, where $a$ is some constant.  The
relative strength of the $\tau^{-\sigma}$ and $\tau^{-2\sigma}$
singularities in the function $L(a,\chi)$ is determined by the value
of $a$. Therefore it might be numerically convenient to analyze the
function $L(a,\chi)$ with $a \simeq -c^{sq}_2$, rather than simply
$\rm ln^2(\chi)$, because the dominant singularity of
$L(-c^{sq}_2,\chi)$ should then be ${\tau}^{-2\sigma}$ , with a small
or vanishing ${\tau}^{-\sigma}$ contribution and thus approximately a
simple pole, if $\sigma \simeq1/2$.  We can therefore expect that
analyzing the DAs of $L(-c^{sq}_2,\chi)$ will make it possible to
determine with good accuracy not only the position, (which is not very
sensitive to the choice of $a$), but also the exponent of the critical
singularity.

 In our analysis of the HT series, we have restricted to a class of
quasi-diagonal second-order DAs quite similar to that used in
Ref.[\onlinecite{bper}], namely to the $[k,l,m;n]$ DAs defined by the
conditions: $20 \leq k+l+m+n \leq 24$ with $k \geq 6; l \geq 6; m \geq5$ 
and such that $|k-l|, |l-m| \leq 2$, and $1 < n < 7$.  For the
shorter {\it tr} lattice series, these conditions have to be modified
in an obvious way.  We have always made sure that our final estimates,
within a small fraction of their spread, are independent of the
precise definition of the DA class examined.

\begin{figure}[tbp]
\begin{center}
\leavevmode
\includegraphics[width=3.37 in]{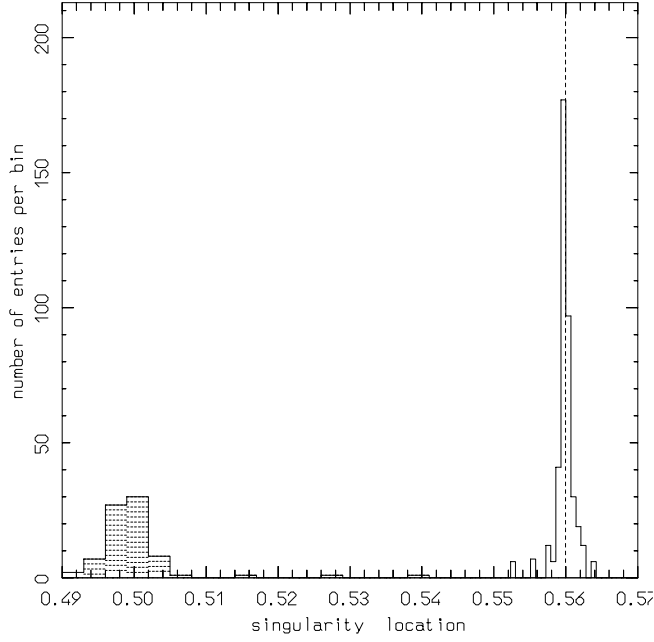}
\caption{\label{figbetacda2}XY model on the {\it sq} lattice.
 Distribution of the singularities of a class
 of second-order DAs of $\rm ln^2(\chi)$ vs their position on the
 $\beta$ axis (open histogram).  The central value of the open
 histogram is $\beta_c=0.5599(7)$. The bin width is 0.0007.  The
 vertical dashed line shows the critical value $\beta_c=0.55995$
 indicated by the simulation of Ref.[\onlinecite{Has}], for which one can guess
 an uncertainty  somewhat smaller than ours.  The hatched
 histogram represents the distribution of the exponent $\sigma$
 obtained from DAs of $L(-c^{sq}_2,\chi)$, biased with $\beta_c= 0.55995$,
 vs  their position on the $\sigma$ axis.  The central value of the
 hatched histogram is $\sigma=0.499(5)$ and the bin width is 0.003.
 }
\end{center}
\end{figure}

In the case of the {\it sq} lattice, from an unbiased DA analysis of
$\rm ln^2(\chi)$ we obtain $\beta_c=0.5599(7)$, in complete agreement
with the results of our previous analysis of the 24-th order series
and also with the MonteCarlo analysis of Ref.[\onlinecite{Has}] which
yielded $\beta_c=0.55995(15)$. Notice that the uncertainty of this
estimate was not given explicitly in Ref.[\onlinecite{Has}], and that
the value that we have nevertheless indicated is only a reasonable
guess. We have obtained the estimate $\sigma=0.499(5)$ for the
singularity exponent by DAs of $L(-c^{sq}_2,\chi)$, chosen in the same
class defined above and biased with the value of $\beta_c=0.55995$
given in Ref.[\onlinecite{Has}].  Its uncertainty accounts also for
the spread in the estimate of the critical temperature used to
bias the DA calculation. These results are illustrated 
in Fig.\ref{figbetacda2}. 

\begin{figure}[tbp]
\begin{center}
\leavevmode
\includegraphics[width=3.37 in]{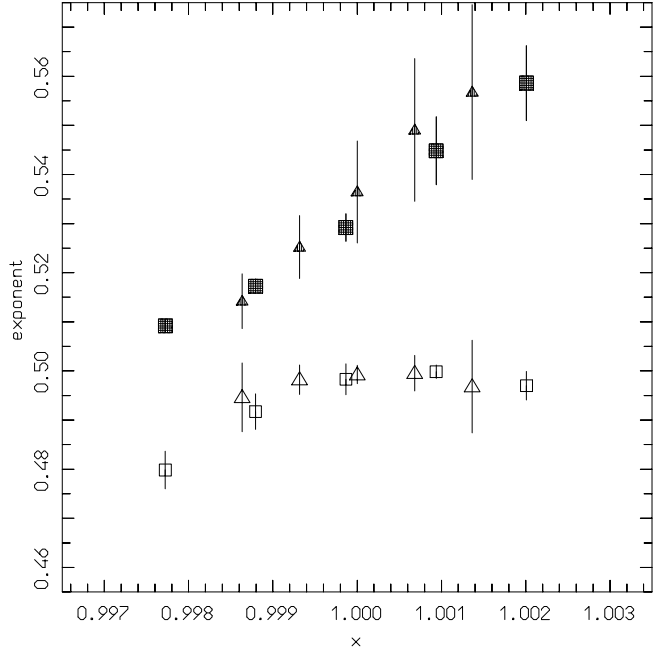}
\caption{\label{figsigma} A comparison of the estimates of the
exponent $\sigma$ obtained from a class of DAs of $\rm ln^2(\chi)$ and
of $L(-c_2,\chi)$, biased with the inverse critical temperature.
Results are shown for both the {\it sq} and the {\it tr} lattices.  We
have varied the value of $\beta^{bias}_c$, used to bias the DAs, in a
small vicinity of the values $\beta^{sq}_c=0.55995$, in the case of the
{\it sq} lattice, and $\beta^{tr}_c=0.3412$, in the case of the {\it tr}
lattice.  The temperature-biased exponent estimates are plotted vs
$x=\beta^{bias}_c/\beta^{sq}_c$ in the case of the {\it sq} lattice and
vs $x=\beta^{bias}_c/\beta^{tr}_c$ in the case of the {\it tr}
lattice.  The black squares (resp. black triangles) show the results
obtained from the analysis of $\rm ln^2(\chi)$ in the case of the {\it sq}
lattice (resp. {\it tr} lattice) and the open squares (resp. open
triangles) those obtained from the study of $L(-c_2,\chi)$ 
in the case of the {\it sq} lattice (resp. {\it tr} lattice).  }
\end{center}
\end{figure}

  In Fig.\ref{figsigma}, we have compared how the estimate of $\sigma$
 depends on the value $\beta^{bias}_c$ used to bias the DAs, when
 either the quantity $L(-c^{sq}_2,\chi)$ or $\rm ln^2(\chi)$ is
 analyzed in a vicinity of $\beta^{bias}_c = \beta_c=0.55995$.
 Clearly the results obtained from the analysis of $\rm ln^2(\chi)$
 are much more sensitive to the bias value.

\begin{figure}[tbp]
\begin{center}
\leavevmode
\includegraphics[width=3.37 in]{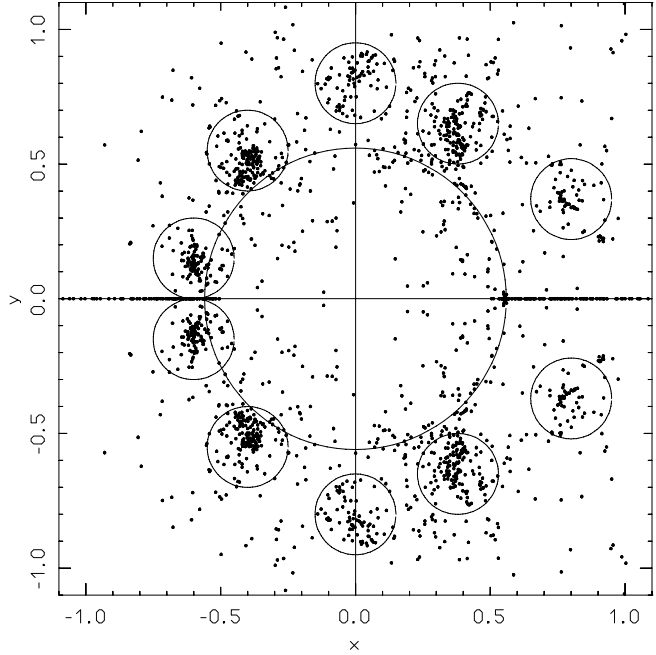}
\caption{\label{figsing} XY model on the {\it sq} lattice.  A scatter
 plot of the singularities of a class of first-order DAs for 
$L(-c^{sq}_2,\chi)$ in the complex $\beta$ plane.  Here $x= Re(\beta)$ and
 $y= Im(\beta)$. The central circle has radius $\beta_c$.  The small
 circles are drawn to enclose clusters of singularities which are
 likely to coalesce around tips of cuts.}
\end{center}
\end{figure}

Taking advantage of our extension of the HT series, we can also get
some hint of the analytic structure of the susceptibility in a
vicinity of the origin of the complex $\beta$ plane.  In
Fig.\ref{figsing}, we have reported a scatter-plot showing the
distribution of the nearby complex singularities for a large class of
inhomogeneous first-order DAs of $L(-c^{sq}_2,\chi)$. We have
discarded only a few spurious real singularities with modulus less
than $0.9\beta_c$.  We can notice that, in addition to the physical
singularities at $\pm \beta_c$ and to many randomly scattered complex
singularities, which we believe to represent only numerical noise,
there is evidence of five pairs of complex-conjugate singularities,
located just beyond the convergence disc of the series. They are
indicated by clusters of DA singularities, which are likely to
coalesce around tips of cuts with increasing series order.  These
results, which do not essentially depend on the value of the constant $a$ in
$L(a,\chi)$, are reminiscent of the regular pattern of the nearby
unphysical singularities (cuts) which were exactly determined for the
{\it sq}-lattice two-dimensional $O(N)$ $\sigma$-model\cite{oinf} in
the large $N$ limit and conjectured\cite{oenne2d} to exist also for
finite $ N\geq 3$.

\begin{figure}[tbp]
\begin{center}
\leavevmode
\includegraphics[width=3.37 in]{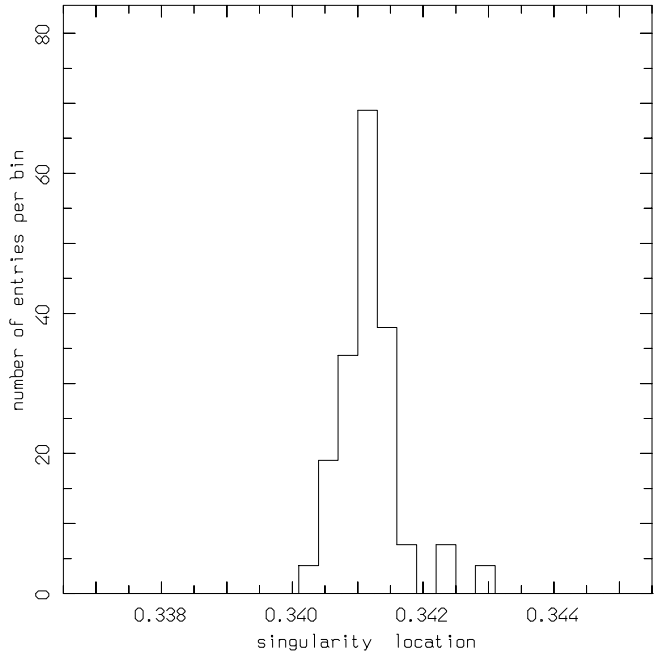}
\caption{\label{figbetactria} XY model on the {\it tr} lattice.
 Distribution of the singularities of a class of second-order
 inhomogeneous DAs of $\rm ln^2(\chi)$ versus their position on the
 $\beta$ axis.  The central value of the distribution
 is $\beta_c=0.3412(4)$. The bin width is 0.0003.  }
\end{center}
\end{figure}

 Let us now perform a similar update of the analysis for the {\it tr}
 lattice series. As shown
 in Fig.\ref{figbetactria},  an unbiased DA study of $\rm ln^2(\chi)$ yields a
 critical inverse-temperature $\beta_c=0.3412(4)$,  sizably improving
 the precision of our 14th-order estimate $\beta_c=0.340(1)$ reported
 in Ref.[\onlinecite{bct}], but disagreeing with the DA estimate
 $\beta_c=0.33986(4)$ obtained in the 15th-order study of
 Ref.[\onlinecite{pisa}].  From an analysis of DAs of $\rm ln^2(\chi)$
 biased with the critical temperature $\beta_c=0.3412(4)$, we are led
 to the estimate $\sigma=0.53(2)$.  If, however, in analogy with the
 {\it sq}-lattice analysis, we consider the quantity $L(a,\chi)$ with
 $a=-c^{tr}_2 \simeq 1.1$, obtained from a fit of the asymptotic
 behavior of $\rm ln(\chi)$ as indicated above, we arrive at the
 estimate $\sigma=0.500(4)$.  It is also particularly interesting to
 notice that the pattern of the singularities of $L(a,\chi)$ in the
 complex $\beta$-plane is much simpler than that observed in the case
 of the {\it sq} lattice and consists of a single pair of
 complex-conjugate clusters, located farther from the border of the
 convergence disc than in the case of the {\it sq} lattice.  This is
 an indication that the convergence of the {\it tr}-lattice HT series
 will be faster than in the {\it sq} lattice case, and so the argument
 that the {\it effective lengths} of the HT series for the two
 lattices are comparable, receives further support.

 In spite of the sizable extension of the
 series, both in the {\it sq}  and in the {\it tr} lattice cases,
 the study of the indicator function $H(\beta)={\rm
 ln}(1+m^{(2)}/\chi^2)/\rm ln(\chi) = \frac{\eta} {2-\eta} +O(\tau^{\sigma})$ 
(or of analogous functions of different correlation-moments), 
does not show  very sharp improvements in the
 accuracy of the determination of the exponent $\eta$, for which we 
 consistently obtain the estimate $\eta=0.25(2)$.

Assuming $\sigma=1/2$, essentially the same estimate of $\eta$
 is obtained from a DA analysis of the series expansions of 
$\tau^{\sigma}\rm ln(\chi)= (2-\eta) {\it b} +O(\tau^{\sigma})$
 and of $\tau^{\sigma}\rm ln(\xi^2/\beta))=2 {\it b} +O(\tau^{\sigma})$ both
 in the case of the {\it sq} and of the {\it tr} lattices.
 Assuming  $\eta=1/4$, we can estimate the non-universal parameters
 $b_{sq}=1.77(1)$ and $b_{tr}=1.70(1)$.

In order to get alternative estimates of $\eta$, we may also try to
 analyze directly the large-order behavior of the expansion
 coefficients of $\chi(\beta)$ and of $\xi^2(\beta)$.  Assuming
 $\sigma=1/2$ and using the known value of $\beta_c$, we can fit the
 series coefficients of these quantities to their expected\cite{BC}
 asymptotic behavior $c_n \sim \beta^{-n-1}
 \exp[B(n+1)^{1/3}+O(n^{-1/3})]$ with 
$B=B_{\chi}=\frac{{3/2} ((2-\eta)b)^{2/3}} {(\beta_c/2)^{1/3}}$ 
in the case of the susceptibility, while 
$B=B_{\xi^2}=\frac{ 3/2 (2b)^{2/3}} {(\beta_c/2)^{1/3}}$ 
in the case of $\xi^2(\beta)$. Both in the case
 of the {\it sq} and of the {\it tr} lattices, using the sets of
 coefficients of the susceptibility and of the correlation length, we
 can estimate $\eta=0.25(1)$ from the ratio $B_{\chi}/B_{\xi^2}$. This
 estimate is however suspect because, assuming $\eta=0.25$, we
 can consistently estimate $b=1.46(10)$ in the case of the {\it sq} lattice
 and $b=1.36(10)$ in the case of the {\it tr} lattice. Thus we must
 suppose that the series coefficients are not yet near enough to their
 asymptotic values that the value of $b$ can be reliably estimated by
 this direct analysis, in spite of the fact that the ratio
 $B_{\chi}/B_{\xi^2}$ already takes the expected value.

 Finally, it is also interesting to notice that, as indicated in our
 previous analysis\cite{bper}, a study of the function
 $R(\beta)=\chi(\beta)/\xi^{2-\eta}(\beta)$, of its logarithm and of
 its log-derivative, supports the arguments of Ref.[\onlinecite{Bal}]
 concerning the absence of singular multiplicative corrections by
 powers of $\rm ln(\xi)$ to the critical behavior of $\chi$ and gives
 results completely consistent with the assumption that
 $\eta=1/4$. Indeed if $\eta \not= 1/4$ or if $\eta =1/4$, but there
 were in $\chi$ singular multiplicative corrections, either Pad\'e
 approximants or DAs of the above quantities should detect some
 singular behavior as $\beta \rightarrow \beta_c$, which, at the
 present orders of expansion, is not seen at all.

In conclusion, our recent HT series data confirm  the BKT
predictions for the critical behavior of the XY system and confirm or
improve the existing estimates of the critical parameters for both the
{\it sq} and the {\it tr} lattices.

\section{Acknowledgements} We thank  H. Arisue  for informing us 
 about his recent results. Our computations have been performed on the pc
 cluster {\it Turing} of the Milano-Bicocca INFN Section. We thank the
 Physics Depts. of Milano-Bicocca University and of Milano University
 for their hospitality and support. Our work was partially supported
 by the  MIUR.

\begin{table}
\caption{ 
 XY model on the {\it tr}  lattice. 
The series expansion coefficients for the  nearest-neighbor spin-spin 
correlation $C(0,0;1,0;\beta)$, 
the reduced susceptibility $\chi(\beta)$ and the second moment
 of the correlation function $\rm m^{(2)}(\beta)$.} 
 \center
\begin{tabular}{|c|c|c|c|}
 \hline
order&$C(0,0;1,0;\beta)$  &  $\chi(\beta) $ & $\rm m^{(2)}(\beta)$  \\
\hline
0 &0  & 1 &0  \\
1 &1  &6  & 6 \\
2 &2  &30  &72  \\
3 &$\frac { 7 } { 2 }$  &135  & 579 \\
4 & 5 & 570 & 3834 \\
5 & $\frac { 35 } { 6 }$ & 2306  & 22520 \\
6 & $\frac { 14 } { 3 }$ & $\frac {18083} {2}$ &121754  \\
7 &$\frac { -81 } { 16 }$  &$\frac {276657} {8}$  &$\frac{4952033}{8}$  \\
8 &$\frac { -3769 } { 72 }$  & $\frac{777805}{6}$  &$\frac{36001013}{12}$  \\
9 &$ \frac { -165161 } { 720 }$ & $\frac {14339641}{30}$ 
 &$\frac{839474407}{60}$  \\
10 &$\frac { -7821 } { 10 }$  &$\frac {208590287}{120}$ 
 &$\frac{ 1264157753}{20}$ \\
11 &$\frac { -20160371 } { 8640 }$  &$\frac{8995595389} {1440}$ 
 &$\frac{ 400323755461}{1440}$ \\
12 &$\frac { -27984359 } { 4320 }$  &$\frac{3199713875}{144}$  
 &$\frac{ 860379412817}{720}$ \\
13 &$\frac { -87289819 } { 5040 }$  &$\frac{65793037351} {840}$ 
  &$\frac{ 12688548065393}{2520}$ \\
14 &$\frac { -10256893919 } { 226800 } $  &$\frac {165647319078571} {604800}$
  &$\frac{3152231835174739}{151200}$  \\
15 &$\frac { -3357272555039 } { 29030400 } $
  &$\frac {4600845479023849} {4838400}$ 
 &$\frac{ 411232110524237321}{4838400}$ \\
16 &$\frac { -1400375733941 } { 4838400 } $
  &$\frac {1983863997387623} {604800}$
  &$\frac{ 826713976281365323 }{2419200}$\\
17 &$\frac { -26431737035251 } { 37324800 } $ 
 &$\frac {24492996075345043} {2177280}$ 
 &$\frac{4220638202244829129 }{3110400}$ \\
18 &$\frac { -55206137402197 } { 32659200 } $ 
 &$\frac {1671043059049640293} {43545600}$&
 $\frac{662836750489256935}{124416}$  \\
19 &$\frac { -6827447251427903 } { 1741824000 } $ 
&$\frac{37817672635562705657} {290304000}$  
 &$\frac{2575741048252255298333}{124416000}$  \\
20 &$\frac { -480824970393676609 } { 54867456000 } $  
&$\frac {4025832361031298767249} {9144576000}$ 
 &$\frac{728769389306358221619671}{9144576000}$ \\ 
\colrule  
\end{tabular} 
\label{tabtria}
\end{table}

\end{document}